\documentclass{elsart}
\usepackage{color,setspace,times}
\usepackage{amssymb,amsmath,graphicx,color,rotating,subfigure,url}
\usepackage[square,sort&compress,comma]{natbib}
\usepackage{lineno}

\journal{Physica A} 

\begin{document}

\begin{frontmatter}

\title{Empirical distributions of Chinese stock returns at different microscopic timescales}%
\author[SB,SS]{Gao-Feng Gu},
\author[SZSE]{Wei Chen},
\author[SB,SS,CES,RCSE]{Wei-Xing Zhou\corauthref{cor}}
\corauth[cor]{Corresponding author. Address: 130 Meilong Road, P.O.
Box 114, School of Business, East China University of Science and
Technology, Shanghai 200237, China, Phone: +86 21 64253634, Fax: +86
21 64253152.}
\ead{wxzhou@ecust.edu.cn} %

\address[SB]{School of Business, East China University of Science and Technology, Shanghai 200237, P. R. China}
\address[SS]{School of Science, East China University of Science and Technology, Shanghai 200237, P. R. China}
\address[SZSE]{Shenzhen Stock Exchange, 5045 Shennan East Road, Shenzhen 518010, P. R. China}
\address[CES]{Center for Econophysics Studies, East China University of Science and Technology, Shanghai 200237, P. R. China}
\address[RCSE]{Research Center of Systems Engineering, East China University of Science and Technology, Shanghai 200237, P. R. China}

\begin{abstract}
We study the distributions of event-time returns and clock-time
returns at different microscopic timescales using
ultra-high-frequency data extracted from the limit-order books of 23
stocks traded in the Chinese stock market in 2003. We find that the
returns at the one-trade timescale obey the inverse cubic law. For
larger timescales (2-32 trades and 1-5 minutes), the returns follow
the Student distribution with power-law tails. With the decrease of
timescale, the tail becomes fatter, which is consistent with the
vibrational theory.
\end{abstract}

\begin{keyword}
Econophysics; Probability distribution; Chinese stocks;
Ultra-high-frequency data; Order book and order flow; Inverse cubic
law; Power-law tail

\PACS 89.65.Gh, 89.75.Da, 02.50.-r
\end{keyword}

\end{frontmatter}

\section{Introduction}
\label{s1:itd}

The distribution of asset price fluctuations has crucial implication
on asset pricing and risk management
\cite{Mantegna-Stanley-2000,Bouchaud-Potters-2000,Malevergne-Sornette-2006}.
In the seminal paper for option pricing, Black and Scholes assume
that asset prices follow geometric Brownian motion, that is, the
returns are normally distributed \cite{Black-Scholes-1973-JPE}. It
is well-known that, for most financial assets, this assumption is
merely a rude approximation at large time scale. In addition, Fama
finds that the portfolio selection in a stable Paretian market is
different from that in a Gaussian market \cite{Fama-1965-MS}. It is
natural that the distribution of returns remains a hot topic
especially when huge databases recording transaction-level time
series of stocks become available, which enables us to test classic
theories and models in finance, such as the mixture of distributions
hypothesis \cite{Clark-1973-Em}.

The modeling of the distribution of price variations in financial
markets can be traced back to the work of Bachelier in 1900
\cite{Bachelier-1900}. Let $S(t)$ denote the price of a security at
time $t$. Bachelier submits that the price variation
\begin{equation}
 \Delta{S}(t)=S(t)-S(t-\Delta{t})
 \label{Eq:DS}
\end{equation}
is an i.i.d. variable and follows Gaussian distribution with zero
mean. As pointed out by Mandelbrot \cite{Mandelbrot-1967-JB}, an
implicit assumption of Bachelier's model is that the variance of
$\Delta{S}(t)$ is independent of the price level $S(t)$ {\em{per
se}}, which however contradicts empirical findings. Nowadays, the
logarithmic return is usually used
\begin{equation}
 r(t)=\ln S(t)- \ln S(t-\Delta{t})~,
 \label{Eq:rt}
\end{equation}
which is a precise approximation of the price growth rate
\cite{Fama-1965-JB}. In addition, investors are more sensitive to
the relative price changes than the absolute changes according to
the Weber-Fechner law \cite{Osborne-1959a-OR}. Quite a few scholars
found evidence supporting the Brownian motion model
\cite{Osborne-1959a-OR,Laurent-1959-OR,Osborne-1959b-OR}. This model
is also called the Bachelier-Osborne model since Osborne
independently rediscovered the model \cite{Fama-1965-JB}.

More than half a century after Bachelier's work, a revolutionary
breakthrough was made by Mandelbrot, who introduced the
Pareto-L{\'{e}}vy distribution to describe the tail of incomes and
speculative price returns
\cite{Mandelbrot-1960-IER,Mandelbrot-1961-Em,Mandelbrot-1963-IER,Mandelbrot-1963-JPE,Mandelbrot-1963-JB}.
The concept of Paretian market is soon accepted by mainstream
financial scholars \cite{Fama-1965-JB}. Using high-frequency data of
the S\&P 500 index, Mantegna and Stanley find that the distribution
of returns can be well characterized by a truncated L{\'{e}vy law
\cite{Mantegna-Stanley-1995-Nature}. Mathematically, the density of
the L{\'{e}vy distribution has a power-law decay in the tail
\begin{equation}
 f(r) \sim r^{-(\alpha+1)}~,
 \label{Eq:PL}
\end{equation}
where $0<\alpha<2$. Pareto finds that the income distribution has a
universal power-law exponent $\alpha=1.5$ \cite{Pareto-1896}, while
Mandelbrot finds that the price fluctuations of cotton give
$\alpha\approx 1.7$ \cite{Mandelbrot-1963-JB}. For the S\&P 500
index, it is found that $\alpha=1.4$
\cite{Mantegna-Stanley-1995-Nature}.

In recent years, new evidence provided by Stanley's group shows that
the tail distributions of many stock indexes and stock prices for
the USA markets exhibit an inverse cubic law
\cite{Gopikrishnan-Meyer-Amaral-Stanley-1998-EPJB,Gopikrishnan-Plerou-Amaral-Meyer-Stanley-1999-PRE,Plerou-Gopikrishnan-Amaral-Meyer-Stanley-1999-PRE},
where the power-law exponent is found to be close to $\alpha=3$. In
contrast, empirical analyses for other stock markets have unveiled
power law tail exponents other than the L{\'{e}}vy regime and the
inverse cubic law. Makowiec and Gnaci\'{n}ski have studied the daily
WIG index (the main index of Warsaw Stock Exchange in Poland) for
five years and found that the distribution of return follows
power-law behaviors in three parts with $\alpha$ equal to 0.76, 2.03
and 3.88 for the positive tail and 0.69, 1.83 and 3.06 for the
negative tail \cite{Makowiec-Gnacinski-2001-APP}. Bertram focuses on
the high-frequency data of 200 most actively traded stocks in the
Australian Stock Exchange in the period from 1993 to 2002, and
reports that the distribution of returns has power-law tails with
$\alpha>3$, which varies with different time interval $\Delta{t}$
from 10 to 60 minutes \cite{Bertram-2004-PA}. Coronel-Brizio {\em et
al}. analyze the daily data (1990-2004) of the Mexican Stock market
index (IPC) and find that the distribution of the daily returns
followed a power-law distribution with the exponent $\alpha^+ =
3.33$ (positive tail) and $\alpha^- = 3.12$ (negative tail) by
selecting a suitable cutoff value \cite{Coronel-Hernandez-2005-PA}.
Yan {\em et al}. investigate the daily returns of 104 stocks (76
from the Shanghai Stock Exchange and 28 from Shenzhen Stock
Exchange) in the Chines stock markets in the period from 1994 to
2001 and argue that the tail exponent is $\alpha^+ = 2.44$ for the
positive part and $\alpha^- = 4.29$ for the negative part
\cite{Yan-Zhang-Zhang-Tang-2005-PA}. After removing the opening and
close returns of high-frequency data for the Shanghai Stock Exchange
Composite index, the tail exponents are much closer to $\alpha=3$
\cite{Zhang-Zhang-Kleinert-2007-PA}.

There are also controversial results for some markets. An example
comes from the Indian stock market. Matia {\em et al}. analyze the
daily returns of 49 largest stocks in the National Stock Exchange
over 8 years (1994-2002) and find that the distribution of daily
returns significantly deviates from the power-law form but decays
exponentially in the form of $P(r) = e^{-\beta r}$ with the decay
coefficient $\beta = 1.34$ for the positive tail and $\beta = 1.51$
for the negative tail \cite{Matia-Pal-Salunkay-Stanley-2004-EPL}. In
contrast, Pan and Sinha have studied the the daily data of two stock
indices (Nifty, 1990-2006 and Sensex, 1991-2006) and found the daily
returns are exponentially distributed followed by power-law decay in
the tails ($\alpha^+ = 3.10$ and $\alpha^- = 3.18$ for Nifty and
$\alpha^+ = 3.33$ and $\alpha^- = 3.45$ for Sensex)
\cite{Pan-Sinha-2007-arXiv}. They also analyze the high-frequency
data of 489 stocks containing the information about all the
transactions carried out in the National Stock Exchange (NSE) for
two-year period (2003-2004) and observe power-law tails with
$\alpha^+ = 2.87$ and $\alpha^- = 2.52$ for $\Delta t = 5$  and
$\alpha \approx 3$ for $\Delta{t}$ ranging from 10 to 60 minutes
\cite{Pan-Sinha-2007-EPL}.

An alternative model for the distribution of returns is the
stretched exponential family, which serves as a bridge between
exponential and power-law distributions
\cite{Laherrere-Sornette-1998-EPJB,Sornette-2004}:
\begin{equation}
 f(r) = \frac{c}{r_0}\left(\frac{r}{r_0}\right)^{c-1} e^{-(r/r_0)^c},
 ~~(r\geqslant0)
 \label{Eq:SE}
\end{equation}
where the distribution approaches to exponential when $c\to1$. The
stretched exponential model has a very interesting behavior when
$c\to0$. If $c(r/r_0)^c\to\beta$ as $c\to0$, then the stretched
exponential density goes to a power law
\cite{Malevergne-Pisarenko-Sornette-2005-QF,Malevergne-Pisarenko-Sornette-2006-AFE}:
\begin{equation}
 f(r) \sim \beta\frac{r_0^\beta}{r^{\beta+1}}~.
 \label{Eq:SE2PL}
\end{equation}
This framework is well verified by empirical analyses
\cite{Malevergne-Pisarenko-Sornette-2005-QF,Malevergne-Pisarenko-Sornette-2006-AFE,Malevergne-Sornette-2006}.

A closely relevant issue concerns with the time scale $\Delta{t}$
defining the return. Roughly speaking, the tail distribution evolves
from power law at small time scale to Gaussian at large scale
\cite{Ghashghaie-Breymann-Peinke-Talkner-Dodge-1996-Nature}, based
on the variational theory in turbulence
\cite{Castaing-Gagne-Hopfinger-1990-PD,Castaing-Gagne-Marchand-1993-PD,Castaing-Chabaud-Hebral-Naert-Peinke-1994-PB,Castaing-1994-PD}.
Numerous empirical studies have been performed in various stock
prices and indexes, such as the S\&P 500 index
\cite{Gopikrishnan-Plerou-Amaral-Meyer-Stanley-1999-PRE,Silva-Prange-Yakovenko-2004-PA,Kiyono-Struzik-Yamamoto-2006-PRL},
the U.S.A. common stocks
\cite{Plerou-Gopikrishnan-Amaral-Meyer-Stanley-1999-PRE}, the Hang
Seng Index for the Hong Kong market \cite{Wang-Hui-2001-JEPB}, and
the KOSPI index and KOSDAQ for the Korean market
\cite{Lee-Lee-2004-JKPS}.

In this work, we utilize a nice database documenting the limit order
flow and individual transactions of 23 Chinese stocks traded on the
Shenzhen Stock Exchange (SZSE). For the Chinese stock market, only
very few efforts were taken to investigate the return distributions
\cite{Yan-Zhang-Zhang-Tang-2005-PA,Zhang-Zhang-Kleinert-2007-PA}. To
the best of our knowledge, there is no literature reporting relevant
results at the transaction level. Our main finding is that the stock
returns obey the inverse cubic law at the transaction level and
thinner power laws at aggregated timescales. The rest of the paper
is organized as follows. In Sec.~\ref{s1:Data}, we describe in brief
the database we use. We investigate the probability distribution of
the event-time returns based on individual trades in
Sec.~\ref{s2:ET} and trade-aggregated returns in Sec.~\ref{s2:ETs}.
The probability distribution of the returns on fixed intervals of
clock time is discussed in Sec.~\ref{s2:CTs}. The last section
concludes.

\section{Data sets}
\label{s1:Data}

The study is based on the data of the limit-order books of 23 liquid
stocks listed on the SZSE  in the whole year 2003. The limit-order
book records ultra-high-frequency data whose time stamps are
accurate to 0.01 second including details of every event. The
tickers of the 23 stocks investigated are the following: 000001
(Shenzhen Development Bank Co. Ltd: 887,741 trades), 000002 (China
Vanke Co. Ltd: 509,360 trades), 000009 (China Baoan Group Co. Ltd:
447,660 trades), 000012 (CSG holding Co. Ltd: 290,148 trades),
000016 (Konka Group Co. Ltd: 188,526 trades), 000021 (Shenzhen Kaifa
Technology Co. Ltd: 411,326 trades), 000024 (China Merchants
Property Development Co. Ltd: 133,586 trades), 000027 (Shenzhen
Energy Investment Co. Ltd: 313,057 trades), 000063 (ZTE Corporation,
265,450 trades), 000066 (Great Wall Technology Co. Ltd: 277,262
trades), 000088 (Shenzhen Yan Tian Port Holdings Co. Ltd: 97,195
trades), 000089 (Shenzhen Airport Co. Ltd: 189,117 trades), 000406
(Sinopec Shengli Oil Field Dynamic Group Co. Ltd: 271,389 trades),
000429 (Jiangxi Ganyue Expressway Co. Ltd: 117,424 trades), 000488
(Shandong Chenming Paper Group Co. Ltd: 120,097 trades), 000539
(Guangdong Electric Power Development Co. Ltd: 114,721 trades),
000541 (Foshan Electrical and Lighting Co. Ltd: 68,737 trades),
000550 (Jiangling Motors Co. Ltd: 346,176 trades), 000581 (Weifu
High-Technology Co. Ltd: 93,947 trades), 000625 (Chongqing Changan
Automobile Co. Ltd: 397,393 trades), 000709 (Tangshan Iron and Steel
Co. Ltd: 207,756 trades), 000720 (Shandong Luneng Taishan Cable Co.
Ltd: 132,233 trades), and 000778 (Xinxing Ductile Iron Pipes Co.
Ltd: 157,321 trades).

In 2003, there are two kinds of auctions on the SESZ, namely the
call auction and continuous double auction. The former refers to the
process of one-time centralized matching of buy and sell orders
accepted during a specified period, while the latter refers to the
process of continuous matching of buy and sell orders on a
one-by-one basis. In each trading day, opening call auction is held
between 9:15a.m. and 9:30a.m., followed by continuous auction
(9:30a.m.-11:30a.m. and 13:00p.m.-15:00p.m.). The orders that are
not executed during opening call auction automatically enter
continuous auction. The limit orders submitted and canceled are
identified by numbers characterizing the aggressiveness and
direction (buyer- versus seller-initiated) of orders. Specifically,
the buyer-initiated (or seller-initiated) orders are differentiated
into six aggressive catalogs from less aggressive to more
aggressive: canceled orders, orders inside the book, orders on the
same best price, orders inside the spread, filled orders, and
unfilled orders
\cite{Biais-Hillion-Spatt-1995-JF,Griffiths-Smith-Turnbull-White-2000-JFE,Hall-Hautsch-2006-EE}.
More information about the market can be found in
Ref.~\cite{Gu-Chen-Zhou-2007-EPJB}.

\section{Empirical distribution of returns}

\subsection{Probability distributions of event-time returns}
\label{s2:ET}

We adopt the midprice of the best bid $b_i(t)$ and best ask $a_i(t)$
of stock $i$ as the price at time $t$ after a transaction occurs:
\begin{equation}
S_i(t) = \frac{b_i(t) + a_i(t)}{2}~,
 \label{Eq:Si}
\end{equation}
where $t$ is the event time corresponding to single trades. Indeed,
the concept of {\em{event time}} was introduced some two score and
odd years ago to study the distribution of returns
\cite{Brada-Ernst-Tassel-1966-OR}. We then define the event-time
return after $\Delta{t}$ trades for stock $i$ as the logarithmic
midprice change:
\begin{equation}
r_i(t) = \ln[S_i(t)/S_i(t - \Delta{t})]~.
 \label{Eq:ri}
\end{equation}
In this section, we focus on $\Delta{t}=1$ trade. In order to treat
all the returns for different stocks as an ensemble, we deal with
standardized returns
\begin{equation}
g_i(t) = \frac{r_i(t)-\mu_i}{\sigma_i}~,
 \label{Eq:gi}
\end{equation}
where $\mu_i$ and $\sigma_i$ are respectively the mean and standard
deviation of returns for stock $i$.

In our analysis, we treat the 23 groups of standardized returns
$g_i(t)$ as an ensemble. The empirical probability density function
$f(g)$ is estimated, as shown on the left panel of
Fig.~\ref{Fig:R:PDF}. We find that $f(g)$ can be well modeled by a
Student density \cite{Blattberg-Gonedes-1974-JB}:
\begin{equation}
f(g|\alpha,m,L) =
\frac{\sqrt{L}\alpha^{\frac{\alpha}{2}}}{B\left(\frac{1}{2},\frac{\alpha}{2}\right)}\left[\alpha
+ L(g-m)^2\right]^{-\frac{\alpha + 1}{2}}~,
 \label{Eq:Student}
\end{equation}
where $\alpha$ is the degrees of freedom parameter (or tail
exponent), $m=\langle{g}\rangle$ is the location parameter, $L$ is
the scale parameter, and $B(a, b)$ is the Beta function, that is,
$B(a, ~b) = \Gamma(a)~\Gamma(b) ~/ ~\Gamma(a + b)$ with
$\Gamma(\cdot)$ being the gamma function. By the definition of
$g_i$, we have $\langle{g_i}\rangle=0$ and thus
$m=\langle{g}\rangle=0$. Nonlinear least-squares regression gives
$\alpha = 3.1$ and $L = 1.9$. The fitted curve is drawn on the left
panel.

\begin{figure}[htb]
\centering
\includegraphics[width=6.5cm]{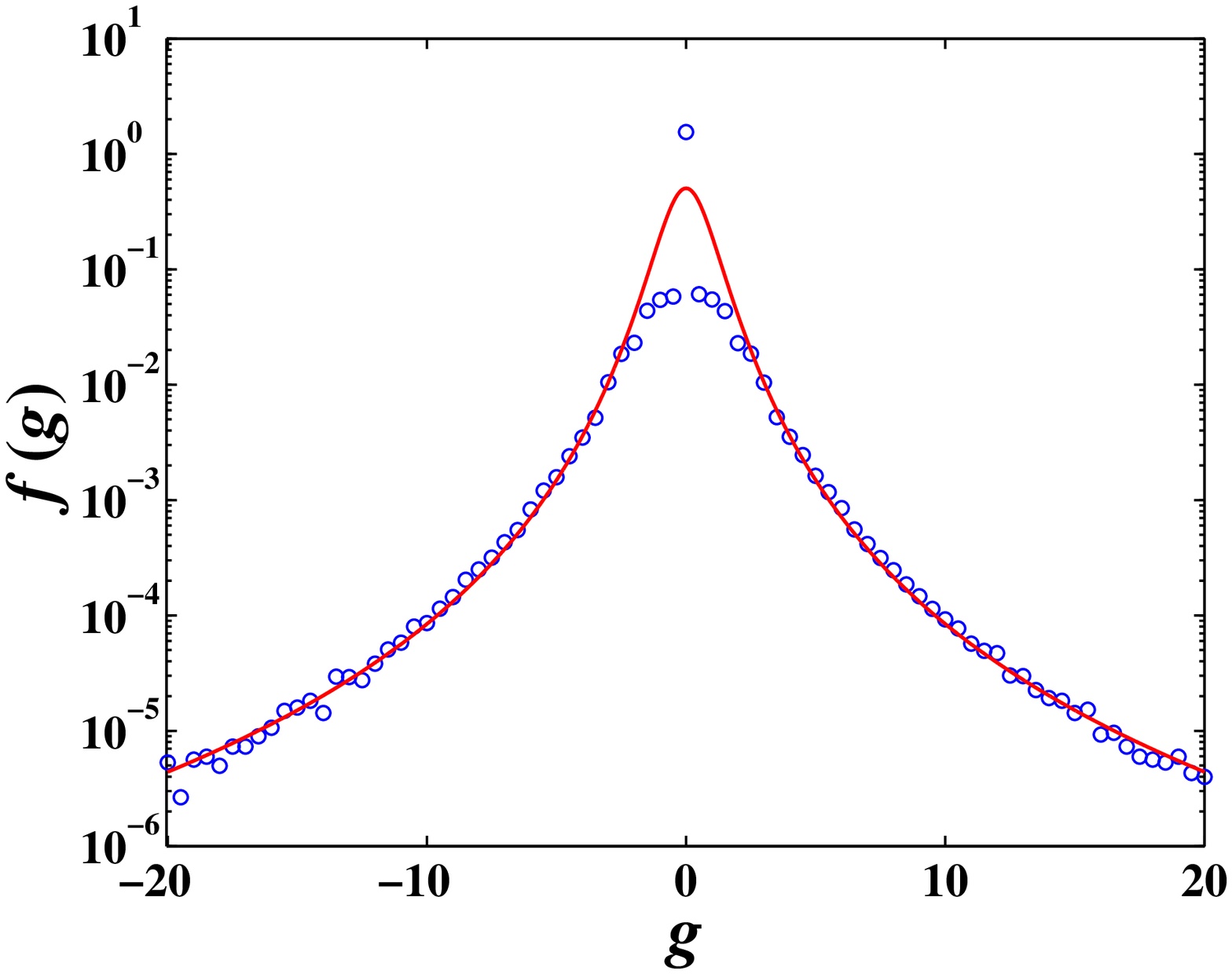}
\includegraphics[width=6.5cm]{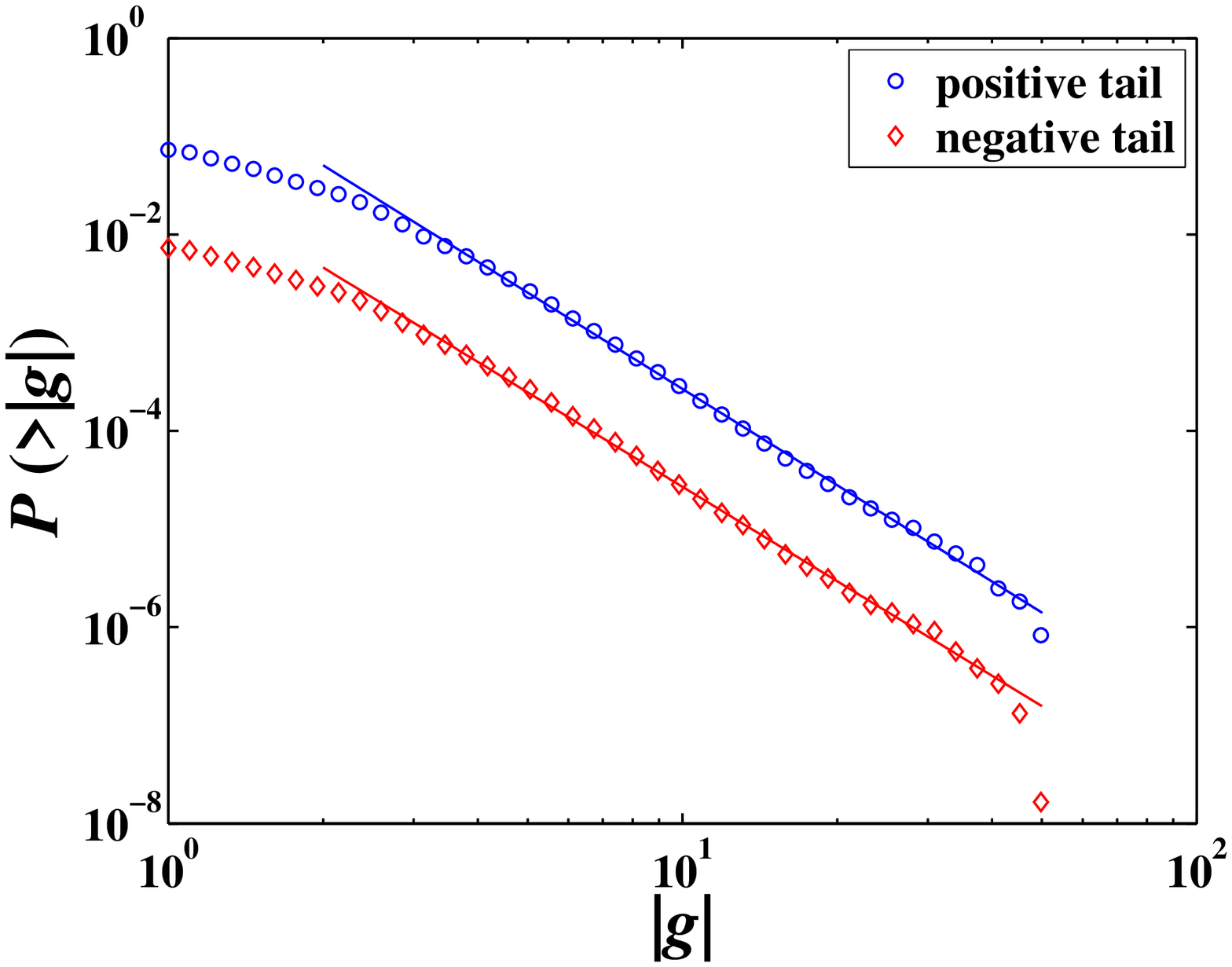}
\caption{\label{Fig:R:PDF}(Color online) Empirical distribution of
event-time returns with $\Delta{t}=1$ transaction. Left: Empirical
probability density function $f(g)$ of the event-time returns $g$
aggregating the 23-stock data. The solid line is the Student density
with $\alpha = 3.1$, $m = 0$ and $L = 1.9$. Right: Empirical
cumulative distributions $P(g)$ for positive and negative normalized
returns $g$. The solid lines are the least squares fits of power
laws to the data with $\alpha^+=3.14 \pm 0.02$ for the positive tail
and $\alpha^-=3.00 \pm 0.02$ for the negative tail.}
\end{figure}

According to the left panel of Fig.~\ref{Fig:R:PDF}, the Student
density fits nicely the tails of the empirical density $f(g)$. The
fitted model deviates the empirical density remarkably for small
values of $|g|$. For large values of $|g|$, the Student density
function $f(g)$ approaches power-law decay in the tails:
\begin{equation}
f\left(g\right) \sim \left\{
 \begin{array}{ccc}
 (-g)^{-(\alpha^- + 1)} && {\rm{for}}~~g<0\\
 (+g)^{-(\alpha^+ + 1)} && {\rm{for}}~~g>0
 \end{array}
 \right..
 \label{Eq:fg}
\end{equation}
The empirical cumulative distributions $P(g)$ of the event-time
returns for positive $g$ and negative $g$ are illustrated in the
right panel of Fig.~\ref{Fig:R:PDF}. Both positive and negative
tails decay in power-law form with $\alpha^+=3.14 \pm 0.02$ and
$\alpha^-=3.00 \pm 0.02$ in line with the tail exponent estimated
from the Student model. We note that the positive and negative tails
are not asymmetric. These results indicate that the standardized
returns obey the inverse cubic law.

\subsection{Probability distributions of aggregated event-time
returns} \label{s2:ETs}

We now turn to investigate the distributions of aggregated
event-time returns, where $\Delta{t}$ spans several trades. By
varying the value of $\Delta{t}$, we are able to compare the PDF's
at different time scales. Specifically, we compare the PDFs for
$\Delta{t}= 2$, $4$, $8$, $16$, $32$ trades with that for
$\Delta{t}=1$ trade. The empirical $f(g)$ functions averaged over 23
stocks for different time scales $\Delta{t}$ are illustrated in
Fig.~\ref{Fig:RNA}(a). We represent the distribution of one-trade
return for comparison. It is evident that the tail is heavier with
the decrease of $\Delta{t}$. This phenomenon can also be
characterized by the kurtosis of the distributions. As listed in
Table~\ref{Tb:RNA}, the kurtosis of each PDF is significantly
greater than that of the Gaussian distribution whose kurtosis is 3,
indicating a much slower decay in the tails. In addition, the
kurtosis decreases with respect to the scale $\Delta{t}$. We also
notice that the PDF for $\Delta{t}=32$ trades decays slower than
exponential.

\begin{figure}[htb]
\centering
\includegraphics[width=6.5cm]{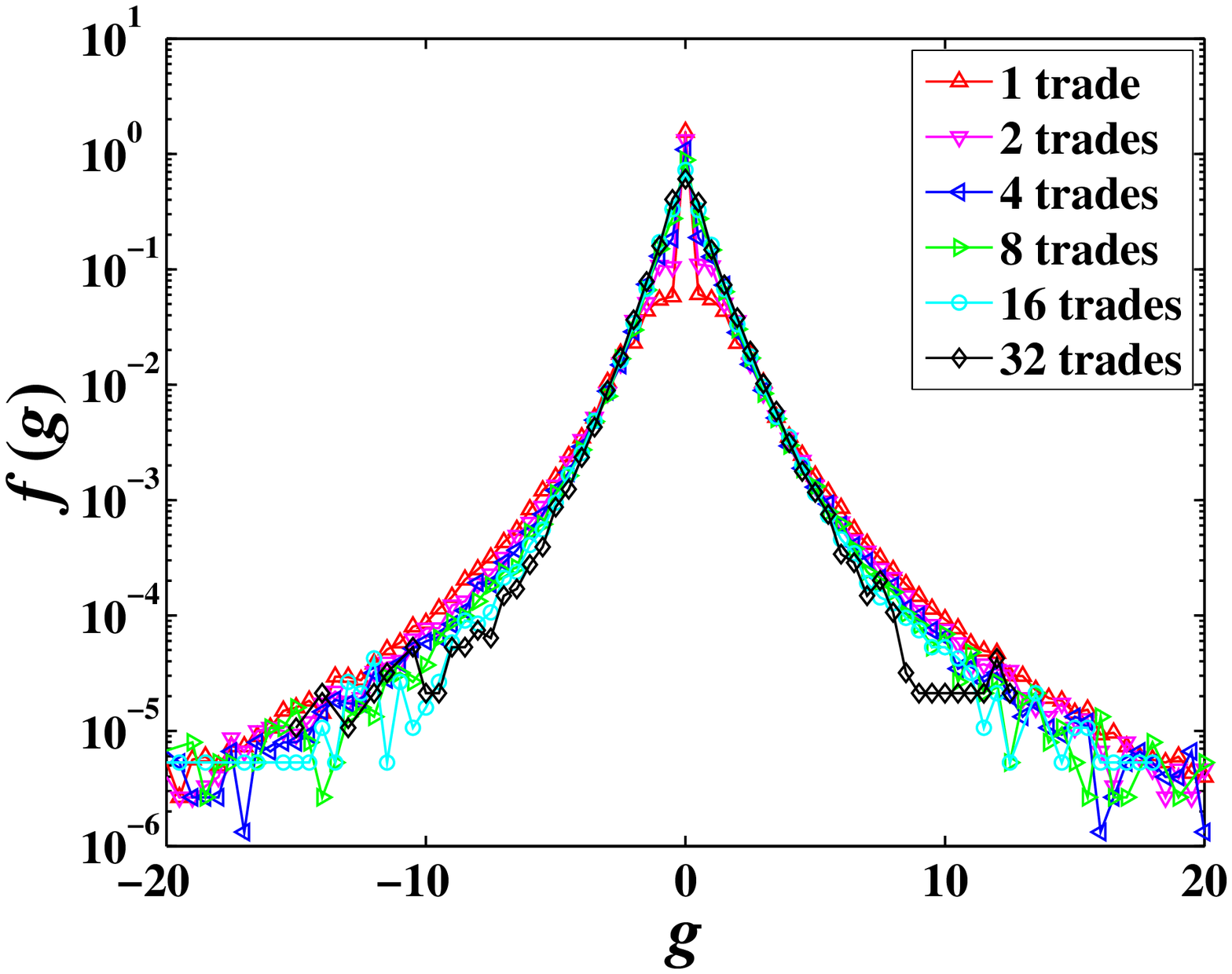}
\includegraphics[width=6.5cm]{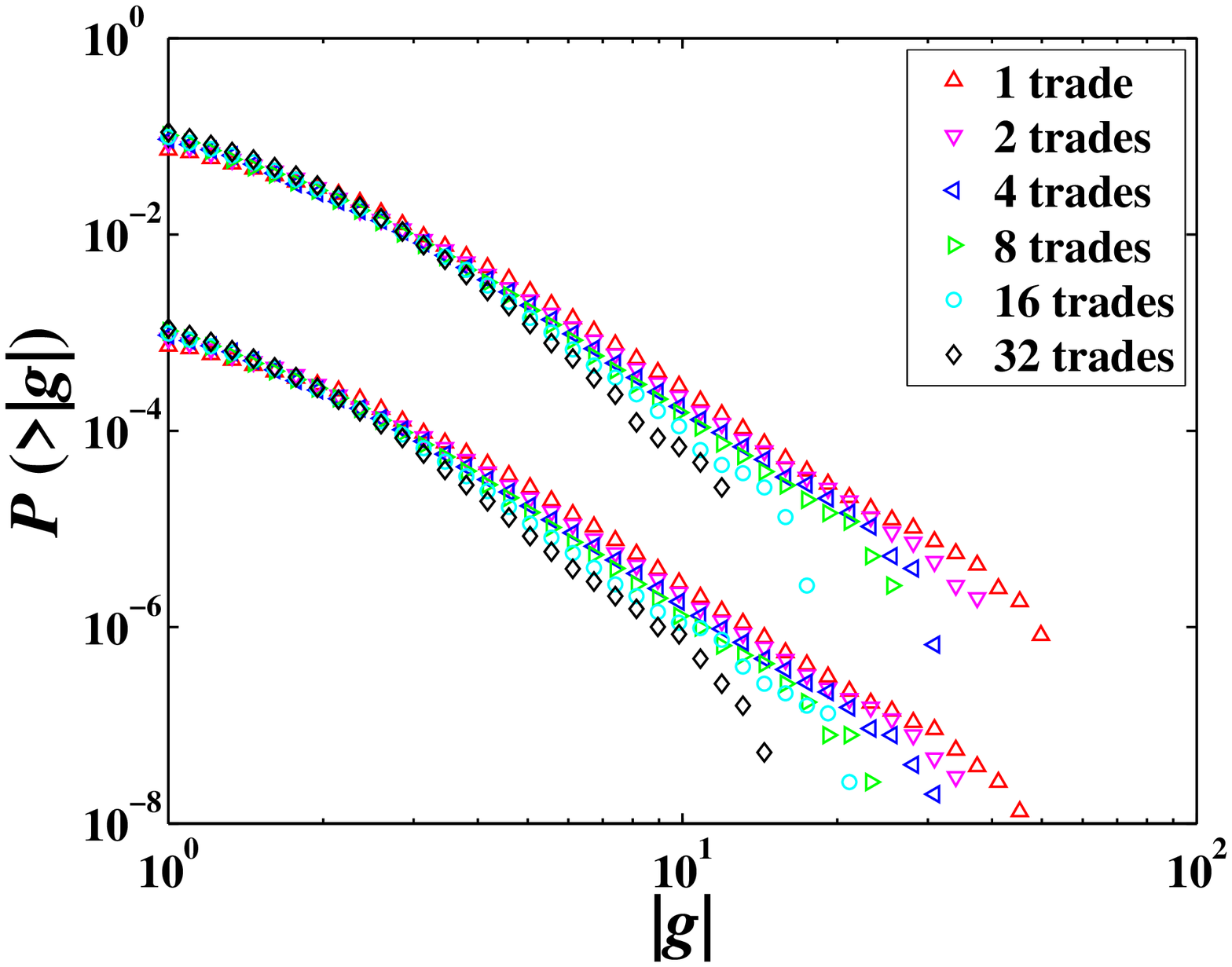}
\caption{\label{Fig:RNA} (Color online) Empirical distributions of
aggregated event-time returns at different time scales $\Delta{t} =
1,~2,~4,~8,~16,~32$ transactions. Panel (a): Empirical densities
$f(g)$ of the aggregated event-time returns. Panel (b): Empirical
cumulative distributions $P(g)$ for positive (upper cluster of
curves) and negative (lower cluster of curves) returns $g$.}
\end{figure}

\begin{sidewaystable}
 \centering
 \caption{Characteristic parameters for aggregated event-time returns.}
 \medskip
 \label{Tb:RNA}
 \centering
 \begin{tabular}{cccccccccccc}
 \hline \hline
    & \multicolumn{2}{@{\extracolsep\fill}c}{Basic statistics} &
    & \multicolumn{2}{@{\extracolsep\fill}c}{Student density} &
    & \multicolumn{2}{@{\extracolsep\fill}c}{Positive tail}  &
    & \multicolumn{2}{@{\extracolsep\fill}c}{Negative tail}  \\
    \cline{2-3} \cline{5-6} \cline{8-9} \cline{11-12}
 $\Delta{t}$ & Skewness & Kurtosis && $L$ & $\alpha$ & &  Scaling range & $\alpha^+$ & &  Scaling range & $\alpha^-$
  \\\hline
  1  & 0.005 & 34.21 && 1.9 & 3.1 && $2.4 \leqslant{g}\leqslant 60.3$ &  $3.14 \pm 0.02$ && $2.1 \leqslant{-g}\leqslant 60.3$ & $3.00 \pm 0.02$\\
  2  & 0.026 & 26.99 && 1.9 & 3.2 && $2.6 \leqslant{g}\leqslant 45.7$ &  $3.18 \pm 0.02$ && $2.4 \leqslant{-g}\leqslant 45.7$ & $3.02 \pm 0.03$\\
  4  & 0.051 & 23.06 && 2.0 & 3.3 && $2.6 \leqslant{g}\leqslant 33.9$ &  $3.33 \pm 0.03$ && $2.4 \leqslant{-g}\leqslant 33.9$ & $3.17 \pm 0.03$\\
  8  & 0.106 & 19.53 && 1.9 & 3.5 && $2.6 \leqslant{g}\leqslant 25.1$ &  $3.63 \pm 0.04$ && $2.4 \leqslant{-g}\leqslant 28.2$ & $3.39 \pm 0.04$\\
  16 & 0.157 & 16.13 && 1.8 & 3.7 && $2.6 \leqslant{g}\leqslant 15.8$ &  $3.87 \pm 0.05$ && $2.4 \leqslant{-g}\leqslant 19.2$ & $3.50 \pm 0.05$\\
  32 & 0.106 & 13.62 && 1.9 & 4.0 && $2.6 \leqslant{g}\leqslant 11.9$ &  $4.11 \pm 0.06$ && $2.6 \leqslant{-g}\leqslant 13.1$ & $3.96 \pm 0.07$\\
  \hline\hline
 \end{tabular}
\end{sidewaystable}

We have fitted the six curves using the Student density model
(\ref{Eq:Student}) and the parameters $L$ and $\alpha$ are listed in
Table~\ref{Tb:RNA}. In Fig.~\ref{Fig:RNA}(b), we study the tail
distributions of the aggregated event-time returns $g$ for different
time scales $\Delta{t} = 2$, $4$, $8$, $16$, $32$ trades. It is
observed that both positive and negative tails follow power-law
distribution. We have estimated the tail exponents, which are
presented in Table~\ref{Tb:RNA}. Note that the scaling range
decreases with increasing $\Delta{t}$, which is also observed for
two Korean indexes \cite{Lee-Lee-2004-JKPS}. As expected, the tails
decay faster for larger $\Delta{t}$, which is consistent with the
behavior of kurtosis. In other words, the tail exponent increases
with $\Delta{t}$, which is validated by Table \ref{Tb:RNA}. It is
interesting to note that the PDF's for large $\Delta{t}$ deviate
significantly from the inverse cubic law. We also find that
$\alpha^-<\alpha<\alpha^+$, which implies that the distributions are
asymmetric and is in line with the positive skewness. It seems that
the sign of $\alpha^--\alpha^+$ is not universal across different
stock markets \cite{Lee-Lee-2004-JKPS}.


\subsection{Probability distributions of clock-time returns}
\label{s2:CTs}

In this section, we handle clock-time returns $r_i(t)$ defined over
a fixed time interval $\Delta t$. The normalized returns $g_i(t)$
are calculated similarly for each stock $i$. Five different time
intervals are selected with $\Delta t = 1$, $2$, $3$, $4$, $5$ min.
The mapping from event time to clock time is nonlinear, which is
determined by the local trading frequency. The trading frequency, as
a measure of stock liquidity, changes from time to time and from
stock to stock. The average trading frequencies per minute for
individual stocks are estimated: 15.74, 4.81, 9.03, 2.08, 7.94,
2.13, 5.14, 2.03, 3.34, 1.22, 7.29, 6.14, 2.37, 1.67, 5.55, 7.05,
4.71, 3.68, 4.92, 2.34, 1.72, 2.79, 3.35. This strongly nonlinear
mapping implies that the distributions of returns in the two
catalogs may behave differently.

Figure~\ref{Fig:RTA}(a) shows the empirical probability density
functions of the returns for $\Delta t = 1$, $2$, $3$, $4$, $5$
minutes. We observe a nice scaling that the five density functions
collapse onto a single curve when $|g|$ is less than 10, which is in
agreement with the fact that the kurtosis listed in
Table~\ref{Tb:RTA} remains unchanged approximately. This scaling is
not surprising. Comparing the kurtosis listed in Table~\ref{Tb:RTA}
with those in Table~\ref{Tb:RNA}, it seems that the five groups of
the clock-time returns are comparable to those aggregated event-time
returns with $\Delta{t}=8$ to 16 trades. Indeed, the tail
distributions differ from each other. We have fitted the five curves
to the Student model (\ref{Eq:Student}) and presented the parameters
$L$ and $\alpha$ in Table~\ref{Tb:RTA}. We find that $\alpha$
increases with $\Delta{t}$, as expected. Note that these
distributions decay faster than the inverse cubic law.

\begin{figure}[htb]
\centering
\includegraphics[width=6.5cm]{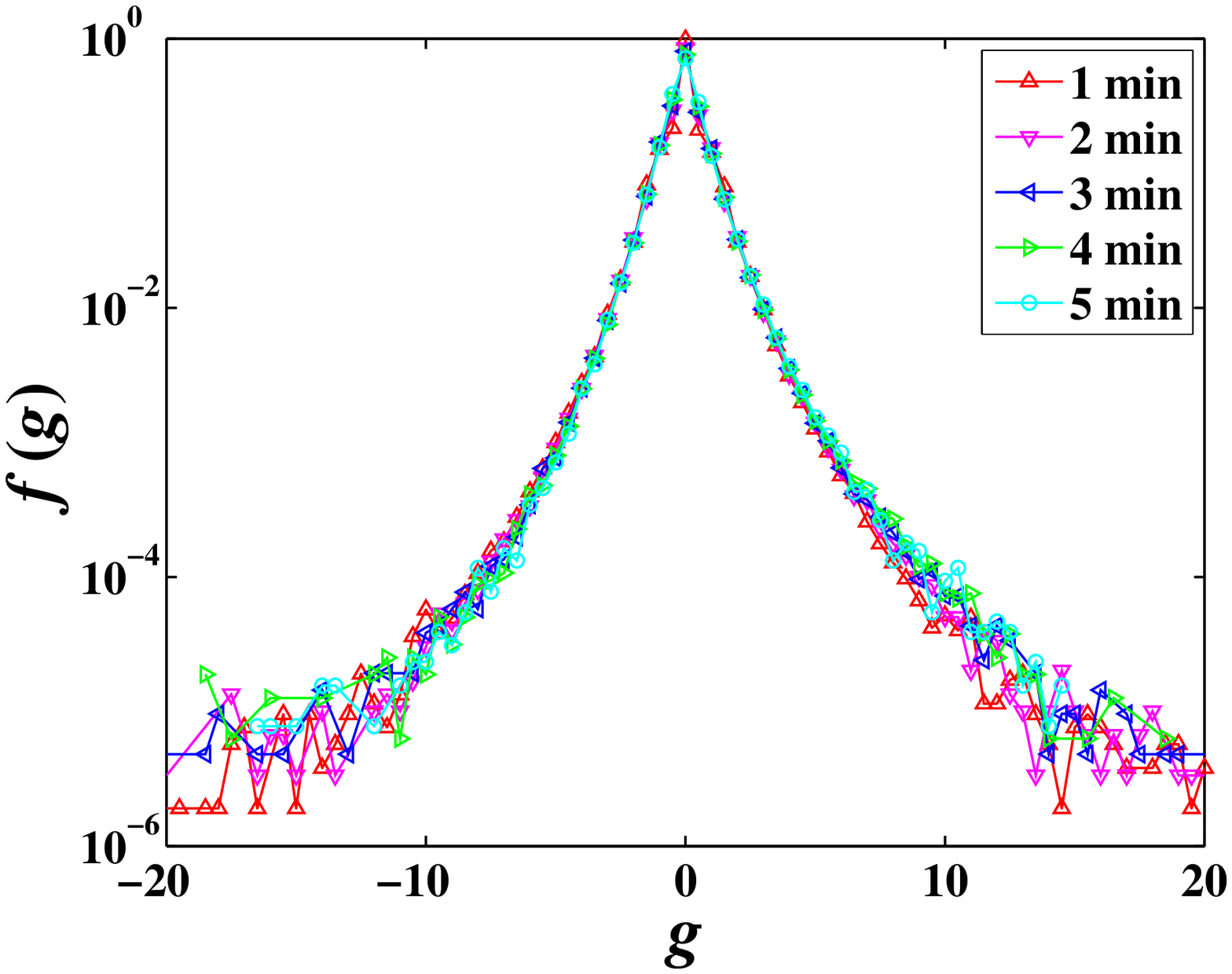}
\includegraphics[width=6.5cm]{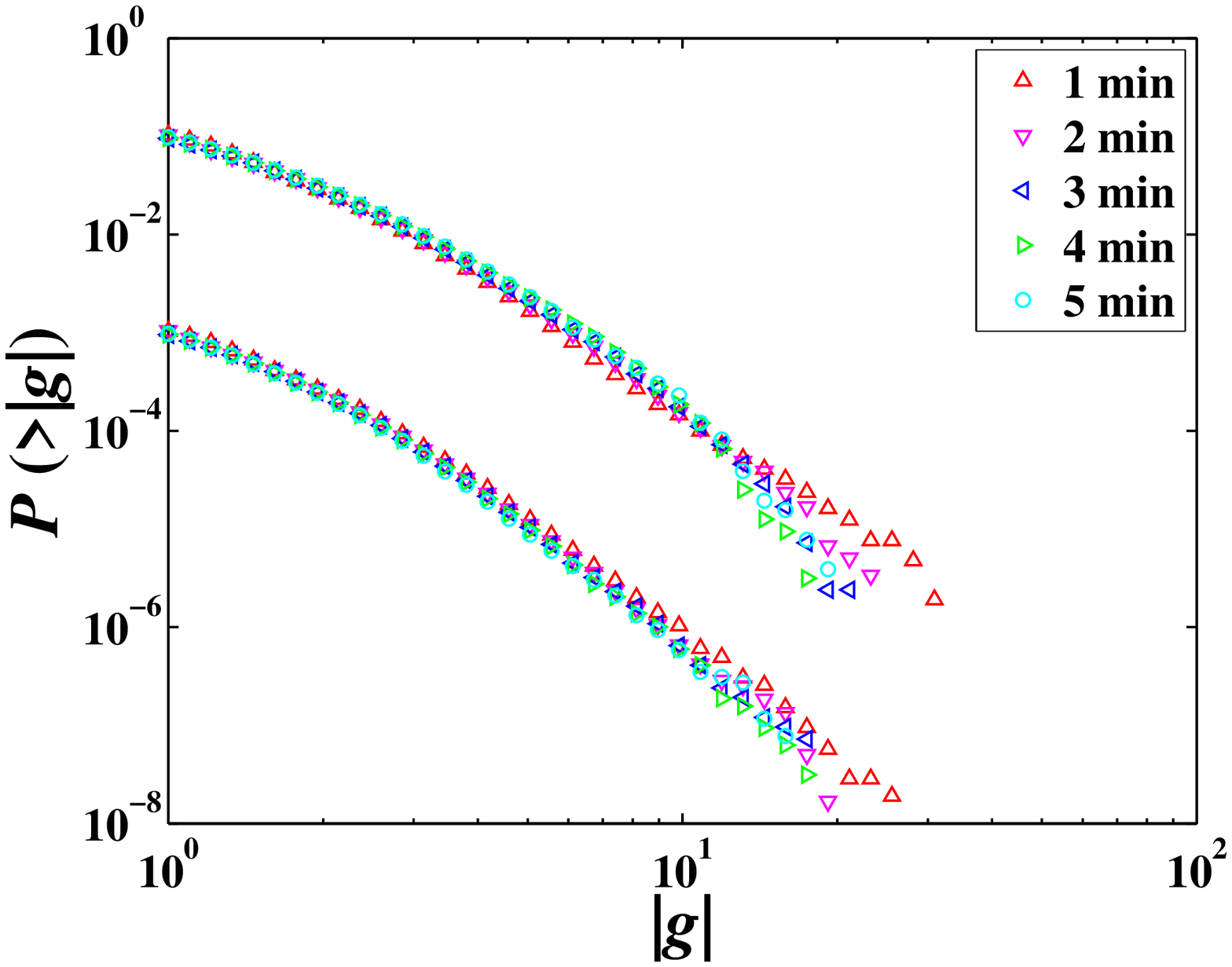}
\caption{\label{Fig:RTA}(Color online) Empirical distributions of
clock-time returns at different time scales $\Delta t =
1,~2,~3,~4,~5$ minutes. Panel (a): Empirical densities $f(g)$ of the
clock-time returns. Panel (b): Empirical cumulative distributions
$P(g)$ for positive (upper cluster of curves) and negative (lower
cluster of curves) normalized returns $g$.}
\end{figure}

We investigate the behavior of tail distributions of the clock-time
returns $g$ for different time scales $\Delta{t} = 1$, $2$, $3$,
$4$, $5$ min in Fig.~\ref{Fig:RTA}(b). We found several similar
characters as those for the aggregated event-time returns and the
reason is possibly that when the time interval becomes large, the
number of trades in this interval increases. Both the positive and
negative tail exponents are estimated, which have been listed in
Table~\ref{Tb:RTA}. Again, the inequality $\alpha^-<\alpha<\alpha^+$
holds roughly.

\begin{sidewaystable}
 \centering
 \caption{Characteristic parameters for clock-time returns.}
 \medskip
 \label{Tb:RTA}
 \begin{tabular}{cccccccccccc}
 \hline \hline
    & \multicolumn{2}{@{\extracolsep\fill}c}{Basic statistics} &
    & \multicolumn{2}{@{\extracolsep\fill}c}{Student densiy} &
    & \multicolumn{2}{@{\extracolsep\fill}c}{Positive tail}  &
    & \multicolumn{2}{@{\extracolsep\fill}c}{Negative tail}  \\
    \cline{2-3} \cline{5-6} \cline{8-9} \cline{11-12}
 $\Delta{t}$ & Skewness & Kurtosis && $L$ & $\alpha$ & &  Scaling range & $\alpha^+$ & &  Scaling range & $\alpha^-$
  \\\hline
  1 & 0.25 & 17.79 && 2.1 & 3.5 && $2.4 \leqslant{g}\leqslant 28.1$ &  $3.48 \pm 0.02$ && $2.1 \leqslant{-g}\leqslant 23.2$ & $3.42 \pm 0.03$\\
  2 & 0.51 & 17.70 && 2.2 & 3.6 && $2.8 \leqslant{g}\leqslant 23.2$ &  $3.70 \pm 0.04$ && $2.1 \leqslant{-g}\leqslant 19.1$ & $3.51 \pm 0.04$\\
  3 & 0.66 & 17.79 && 1.9 & 3.8 && $2.8 \leqslant{g}\leqslant 17.8$ &  $3.92 \pm 0.06$ && $2.4 \leqslant{-g}\leqslant 17.4$ & $3.58 \pm 0.05$\\
  4 & 0.68 & 18.26 && 1.6 & 3.9 && $2.8 \leqslant{g}\leqslant 14.4$ &  $4.05 \pm 0.06$ && $2.4 \leqslant{-g}\leqslant 17.4$ & $3.67 \pm 0.05$\\
  5 & 0.81 & 17.24 && 1.5 & 4.1 && $2.8 \leqslant{g}\leqslant 13.2$ &  $4.22 \pm 0.07$ && $2.4 \leqslant{-g}\leqslant 15.8$ & $3.89 \pm 0.06$\\
  \hline\hline
 \end{tabular}
\end{sidewaystable}

\section{Conclusion}
\label{s6:cns}

Return is among the most important variables in the study of
financial markets and its distribution has crucial implications in
asset pricing and risk management. In this work, we have
investigated a nice database constituting ultra-high-frequency data
extracted from the limit-order books of 23 stocks traded on the
Shenzhen Stock Exchange during the whole year of 2003. We have
studied two types of returns based on event time and clock time,
respectively. We find that the distributions of returns at different
microscopic timescales ($\Delta{t} =1$, $2$, $4$, $8$, $16$, $32$
trades for event-time returns and $\Delta{t} = 1$, $2$, $3$, $4$,
$5$ minutes for clock-time returns) show power-law tails. All the
distributions at different timescales can be well modeled by Student
distributions with different tail exponents. For both types of
returns, the tail exponent increases with the timescale $\Delta{t}$
and the exponent for the positive tail is greater than that for the
negative tail at a fixed timescale. The inverse cubic law is
observed only for the one-trade event-time returns.

\bigskip
{\textbf{Acknowledgments:}}

This work was partly supported by the National Natural Science
Foundation of China (Grant no. 70501011), the Fok Ying Tong
Education Foundation (Grant No. 101086), and the Shanghai
Rising-Star Program (No. 06QA14015).

\bibliography{E:/Papers/Auxiliary/Bibliography}

\end{document}